\documentstyle[aps,preprint,floats,psfig]{revtex}
\begin{document}
\newcommand {\be}{\begin{equation}}
\newcommand {\ee}{\end{equation}}
\newcommand {\bea}{\begin{eqnarray}}
\newcommand {\eea}{\end{eqnarray}}
\newcommand {\nn}{\nonumber}

\draft
%
%
%
%

\title{
Paramagnetic Reentrance Effect in NS Proximity Cylinders
}

\author{Kazumi Maki and Stephan Haas}
\address{Department of Physics and Astronomy, University of Southern
California, Los Angeles, CA 90089-0484}

\date{\today}
\maketitle

\begin{abstract}
A scenario for the 
unusual paramagnetic reentrance behavior
at ultra-low temperatures in Nb-Ag, Nb-Au, and Nb-Cu cylinders
is presented.
For the diamagnetic response down to temperatures of the order 15 mK, 
the standard
theory (quasi-classical approximation) for superconductors appears
to work very well, assuming that Ag, Au, and Cu remain in the normal
state except for the proximity-induced superconductivity. Here it is
proposed that these noble metals may become p-wave superconductors with
a transition temperature of order 10 mK. Below this temperature, p-wave
triplet superconductivity emerges around the periphery of the cylinder.
The diamagnetic current flowing in the periphery is compensated
by a quantized paramagnetic current in the opposite direction, thus
providing a simple explanation for the observed increase in the
susceptibility at ultra-low temperatures.
\end{abstract}

\vskip2pc

In 1990 Visani {\it et al.}\cite{visani}
reported a surprising paramagnetic reentrance phenomenon
at ultra-low temperatures. When a Nb cylinder of diameter $\sim$ 20-100
$\mu$m, covered with a thin film of Ag with a thickness of a few $\mu$m, is
cooled below the superconducting transition temperature of Nb,
the systems initially exhibits the expected diamagnetic response down to
temperatures around 10 mK. However, when the temperature is further 
lowered, the uniform magnetic susceptibility starts to increase again,
indicating a decreasing diamagnetic response at ultra-low temperatures
as $\rm T \rightarrow 0$.
A very similar observation was later reported for an analogous
Nb-Cu system\cite{mota1,muller}, and more recently also for a Nb-Au proximity 
cylinder\cite{mota2}.

The standard
theory for the proximity effect is most conveniently described in terms
of the quasi-classical approximation, which is a more refined version of
the approach first discussed in Ref. \cite{orsay}
Within this approach, one can describe
the diamagnetic response of an S-N system down to 100 mK perfectly well
with only one adjustable parameter, the quasi-particle mean free path
in the normal state\cite{muller,belzig}.

The sudden failure of this quasi-classical approach below 100 mK,
suggested by the observed reentrance behavior, implies that some crucial and
new element is missing from the usual model. Therefore, Bruder and
Imry\cite{bruder} have
proposed a new kind of persistent current around the edge of the
normal metal, circulating in the direction opposite to the diamagnetic
current\cite{bruder}. Although this current is associated with an extended state
in the weak localization theory,
it turns out from a simple estimate that it is of the order of
$10^{-3}$ smaller that the one required to accurately describe the experiments.
More recently, Fauch\`ere {\it et al.}\cite{fauchere}
have proposed that the pairing
interaction in noble metals, such as Cu, Ag, or Au, is repulsive. This implies
that the sign of $\Delta(r)$ changes at the N-S boundary, thus generating
an intrinsic $\pi$-junction at the boundary. As in the model by Bruder and
Imry, this $\pi$-junction could generate a current in the
direction
opposite to the diamagnetic current, resulting in a paramagnetic reentrance
effect
at ultra-low temperatures. However, a Stoner-analysis suggests that
the repulsive interaction used by these authors
is likely to cause a magnetic or charge-density-wave 
instability with a transition temperature of $T^M_c \sim$ 100 mK or
higher. 
Other sets of
experiments\cite{hoyt} on the proximity effect appear to exclude such a large
repulsive potential. Furthermore, if the pairing potential is repulsive,
we would rather expect p-wave superconductivity in these noble metals if
we follow the analysis of Kohn and Luttinger\cite{kohn}.

Let us therefore
propose here that p-wave superconductivity is generated in the outer
film
below a critical temperature of $T_c \sim$ 10 - 100 mK. Earlier
experiments\cite{hoyt} have so far not excluded the possibility of anisotropic
superconductivity in Cu, Ag, or Au at ultra-low temperatures.
The main problem for the observation of these transitions 
is that anisotropic superconductors are highly sensitive to
disorder. For example, assuming a p-wave transition temperature in the regime
of $T_c \simeq 0.1 K$, a sample with a quasi-particle mean free path of
10 $\mu$m or longer would be needed.
In our proposed scenario for the NS proximity cylinders, an additional order 
parameter
$\Delta_p(r)$, associated with intrinsic
p-wave superconductivity in the outer film, has to establish 
itself below $T_c$ against the
presence of the proximity-generated s-wave superconductivity with $\Delta_s(r)$
penetrating into the outer film.
The p-wave superconducting
ordering will thus generate a counter-current, reducing
the kinetic energy associated with $\Delta_p(r)$. 
This counter-current will be quantized, and an approximate expression 
can be derived by minimizing the kinetic energy, as we will show here.

We assume that the London penetration depth of the thin film is larger
than $d_N$, the thickness of the film. The kinetic energy associated
with the p-wave superconductor is then approximately given by 
\begin{equation}
E_{kin} = \frac{1}{2} \rho_S^p (2 e B r - \frac{2\pi n}{l})^2.
\end{equation}
Here $\rho_S^p$ is the superfluid density of the p-wave superconductor,
$n$ is the integer quantum number of the quantized current, and $l$
is the circumference of the thin film, encircling the inner
s-wave superconductor. By minimizing with respect to $n$ we find\cite{haas}
\begin{equation}
n = 2 e B (l/(2\pi))^2 = 0.7958 B l^2,
\end{equation}
where $B$ and $l$ are
expressed in gauss and $\mu$m respectively. Hence it is very likely that
a spontaneous counter-current with $n = 1,2,3, ...$ is generated, compatible
with the actual experimental conditions\cite{visani,mota1,muller,mota2}.
In deriving Eq. 2 we used $r \simeq l/(2\pi)$, and $d_N \ll l$. 

The spatial variation of the magnetic field $B(r)$ is obtained from
\begin{equation}
B_e - B(r) = \frac{1}{\lambda_p^2}  \int_r^{r0} dr' \left( A_{\phi}(r')
- \frac{n}{\phi_0 l} (r_0 - r') \right),
\end{equation}
where $A_{\phi}(r) = \int_0^r dr' B(r')$ is the azimuthal component
of the vector potential.
This leads to a simple differential 
equation,
\begin{equation}
\frac{\partial^2B(r)}{\partial r^2} = \frac{1}{\lambda_p^2} B(r),
\end{equation}
where
$\lambda_p^{-2}= \rho_s^p 4\pi e^2/m$ is the magnetic penetration depth, and
$\rho_s^p$ is the superfluid density of the p-wave superconductor.
The solution
\begin{equation}
B(r) = B_e \exp{[-(r_0-r)/\lambda_p]} + \frac{n}{\lambda_p^2\phi_0 l}
(r - r_0)
\end{equation}
is valid in the outer region $r_0 - r \leq d_N$. 
$B(r)$ is exponentially suppressed in the inner Nb-cylinder, 
as shown in Fig. 1.

\begin{figure}[h]
\centerline{\psfig{figure=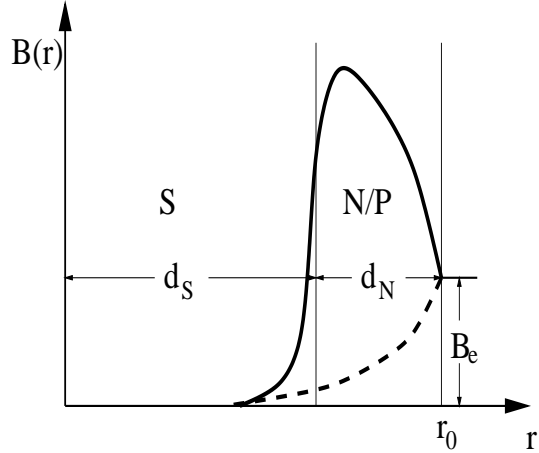,width=7cm,height=6.0cm,angle=0}}
\vspace{0.5cm}
\caption{
Sketch of the spatial dependence of the magnetic field in a normal/s-wave (NS)
or p-wave/s-wave (PS) proximity 
cylinder. If there is p-wave superconductivity in the outer 
layer, the dominant magnetic response is paramagnetic (solid line),
with a maximum at the PS interface. The much smaller 
diamagnetic response is confined to the region close to the surface
of the device.
}
\end{figure}

In recent experiments\cite{visani,mota1,muller,mota2} it was found that the
onset temperature $T^*$ of this paramagnetic
reentrance behavior appears to be inversely proportional to the
length of the cylinder periphery. As shown in Ref.\cite{visani,mota1,muller,mota2},
the experimental data can be represented
by
\begin{equation}
\chi_r(T) = A \exp{(-\frac{T}{T^*})},
\end{equation}
where $A l^2 = const.$ and 
\begin{equation}
T^* = \frac{\hbar v_F}{2\pi k_B l}.
\end{equation}

Let us first attempt to understand the zero-temperature limit of Eq. 6,
i.e. the dependence of the prefactor $A$ on the system dimension
$l$. In the absence of a paramagnetic current, the magnetic field
is almost completely pushed out of the sample except near the edge
of the thin outer film (dashed line in Fig. 1). Once this layer
becomes a p-wave superconductor at ultra-low temperatures, 
the paramagnetic
counter current at the PS interface changes $B(r)$ as indicated by 
the solid line in Fig. 1. Assuming $d_N \ll  d_S$, the magnetic
field in the sample may then be approximately be expressed as
\begin{equation}
\bar{B} \simeq \frac{n\phi_0}{\pi (l/2\pi)^2} = 4\pi n \phi_0 l^{-2},
\end{equation}
where $n$ is a small integer. Under these conditions we can expect that 
the corresponding susceptibility is given by 
\begin{equation}
\chi_r(T=0) = 4\pi n \phi_0 (B l^2)^{-1},
\end{equation}
in agreement with the experiments,
and thus $\chi_r$ diverges as $B^{-1}$. Damping effects 
may smooth out this divergence  as $B \rightarrow
\frac{B}{B_0^2 + B^2}$, consistent with Fig. 3 in Ref. \cite{muller}.

At finite temperatures, it appears that thermal phase fluctuations are not
negligible anymore. Let us first recall that the superfluid density
in the Ginzburg-Landau region is given by $\rho_S^p \sim
|\Delta_p|^2$, where $\Delta_p$ is the superconducting order parameter
of the p-wave superconductor. Since we are considering a quantized
flux around the cylinder, the phase coherence along the periphery 
becomes of crucial importance. Taking into account the possible loss
of phase coherence, let us replace $|\Delta_p|^2$ by
$|\Delta_p|^2 \langle \exp{(i\phi(l) - i\phi(0))} \rangle$ with
\begin{equation}
\langle \exp{(i\phi(l) - i\phi(0))} \rangle =
\exp{[-\frac{1}{2} \langle (\phi(l) -\phi(0))^2 \rangle ] }.
\end{equation}
The average
$\langle (\phi(l) -\phi(0))^2 \rangle$ may be evaluated within the
one-dimensional model along the azimuthal direction of the cylinder
as
\begin{equation}
\langle (\phi(l) -\phi(0))^2 \rangle = \frac{2T}{N(0)} \int
\frac{dq}{2\pi} \frac{1 - \cos{(ql)}}{\xi_0^2 q^2}
\simeq \frac{Tl}{N(0)\xi_0^2 },
\end{equation}
for $T < T_c$  and $\xi_0^2 = \frac{7\xi (3)v_F^2}{2(4\pi T_c)^2}$.
These azimuthal fluctuations along the periphery of the cylinder
destabilize the diamagnetic response in favor of the paramagnetic
counter-current at low temperatures. 

Taking into account the length $L$ of the cylinder, this result can be
substituted into the expression for $\rho_S^p$. It is then 
found that the
superfluid density of the p-wave superconductor reduces to
\begin{equation}
\rho_S^p \rightarrow \rho_S^p\exp{(-\frac{T l L}{N(0)\xi_0^3 })} =
\rho_S^p\exp{(-\frac{T}{T^*})}
\end{equation}
due to the phase fluctuations. Perpendicular fluctuations along the
cylinder are neglected in this context because they play a subdominant
role in stabilizing the counter-current along the PS interface.\cite{footnote} 

Hence we can offer an explanation for the observed $T$- and $l$-dependence
of the exponent, as suggested by the experiments (Eq. 6). In particular,
the above expression for the superfluid density implies that
\begin{equation}
T^* = \frac{N(0)\xi_0^3}{l L} = \frac{m p_F \xi_0^3}{2\pi^2 l L},
\end{equation}
if the density of states $N(0)$ for a 3D system is used. Here $m$
is the quasiparticle mass.
Within this approach, $T^*$ exhibits the observed
$l$-dependence (Eq. 6). However, the numerical value which is obtained 
for $T^*$ is still much larger than 
the experimentally
one, $T^* \approx \frac{v_F}{2\pi k_B l}$. This fact may be remedied 
by considering a 2D density of states $N(0)$ instead, normalized 
by the width $d_N$ of the periphery: $N(0)_{2D} = \frac{m}{2\pi d_N}$. 

In addition, other possible fluctuations should be considered which 
may reduce $T^*$ even further.\cite{kt}
In quasi-one-dimensional systems, phase coherence can be broken by thermal
excitations of vortex pairs or phase slip centers\cite{maki}. If the 
spatial extension of the phase slip centers is of the order of $\xi$
with $\xi = \frac{v_F}{2\pi k_B T}$, it is perhaps plausible to have
a factor $\exp{(-l/\xi )}$, as observed in the experiments, since the
phase slip centers cannot be densely populated. In any case, a
quantitatively correct 
interpretation of the temperature-dependence in the exponential factor
appears to be difficult to find. 

In conclusion, we propose (1) that noble metals may become p-wave
superconductors with $T_c \sim 10 - 100$mK. (2) With this assumption,
the paramagnetic reentrance behavior at ultra-low temperatures can be
described in a quantitative way. (3) Therefore this behavior should
not extend beyond noble metals. More experiments with 
Pt, Ir, and Os would be of great interest.


\begin{references}
\bibitem{footnote} Fully isotropic two-dimensional phase fluctuations lead
to a logarithmic dependence of the superfluid density on the spatial
separation, inconsistent with the experimental data. Hence, we assume that
the dominant phase fluctuations, destabilizing the persistent diamagnetic
current, are in the azimuthal direction.
\end{references}
\end{document}